\documentclass{aa} 

\usepackage{graphicx}

\begin{document}

\title{VLT/NACO infrared adaptive optics images of small scale structures in OMC1
 \thanks{Based on observations collected at 
the ESO/Paranal YEPUN telescope, Proposal 70.C-0315).}}

\author{		F. Lacombe\inst{1} 
	\and 	E. Gendron\inst{1}
	\and 	D. Rouan\inst{1} 
	\and 	Y. Cl\'enet\inst{1}
	\and 	D. Field\inst{2}
	\and 	J.L. Lemaire\inst{3}  
	\and 	M. Gustafsson\inst{2}
	\and 	A.-M. Lagrange \inst{4}
	\and 	D. Mouillet \inst{4}
	\and 	G. Rousset \inst{5}
	\and 	\dag B. Servan \inst{6}
	\and	C. Marlot \inst{1}
	\and       P. Feautrier \inst{4}
	}   
\offprints{daniel.rouan@obspm.fr}  
\institute{
LESIA -- Observatoire de Paris-Meudon, UMR 8109 CNRS, 92195 Meudon, 
France.
\,\, e-mail: francois.lacombe@obspm.fr and daniel.rouan@obspm.fr
\and Institute for Physics and Astronomy,\AA{}rhus University, 8000 \AA{}rhus C, 
Denmark.
\,\, e-mail: dfield@phys.au.dk 
\and LERMA -- Observatoire de Paris-Meudon, UMR 8112 CNRS, 92195 Meudon, France 
and Universit\'e de Cergy-Pontoise, 95031 Cergy Cedex, France.
\,\, e-mail: jean-louis.lemaire@obspm.fr
\and  LAOG -- Observatoire de Grenoble, UMR 5571 CNRS, 38041, Grenoble, France 
\and ONERA -- DOTA,  92322 Chatillon, France 
\and GEPI -- Observatoire de Paris-Meudon, UMR 8111 CNRS,  92195 Meudon,  France 
}

\date{Received 15 March 2003 / Accepted 7 July 2003}

\abstract{
Near-infrared observations of line emission from excited H$_2$
and in the continuum are reported in the direction of the Orion
molecular cloud OMC1, using the European Southern Observatory Very Large
Telescope UT4, equipped with the NAOS adaptive optics system on the
CONICA infrared array camera. Spatial resolution has been achieved at
close to the diffraction limit of the telescope (0.08" -- 0.12" ) and images show a wealth
of morphological detail. Structure is not fractal but shows two
preferred scale sizes of 2.4" (1100 AU) and 1.2" (540 AU), where the larger scale may be
associated with star formation. 
\keywords{ 
ISM: individual objects: OMC1
- ISM: circumstellar matter - ISM: kinematics and dynamics -  ISM:
molecules - Infrared: ISM 
}
}

\titlerunning{OMC1: Small scale structures in OMC1 ...}

\authorrunning{F. Lacombe et al. }

\maketitle   
\section {Introduction}
Near infrared (NIR) observations, performed with the European Southern
Observatory Very Large Telescope (ESO-VLT), are described which provide
high spatial resolution images of the Orion Molecular Cloud (OMC1;
distance 460 pc (Bally et al. 2000)). Data are presented in the v=1-0
S(1) line of H$_2$ at 2.121$\mu$m and in continuum emission at nearby
wavelengths. Earlier NIR imaging of this zone may be found for example
in Allen\&Burton 1993; Brand 1995; Schild et al. 1997;
McCaughrean\&MacLow 1997; Chrysostomou et al. 1997; Chen et al. 1998;
Stolovy et al. 1998; Schultz et al. 1998; Lee\&Burton 1999; Tedds et
al.1999; Salas et al. 1999; Vannier et al. 2001 (V2001); Gustafsson et
al. 2003 (G2003); Kristensen et al.2003 (K2003). The nature of the Orion Nebula Cluster,
in which OMC1 is embedded, is reviewed in O'Dell 2001 and Ferland 2001.
As discussed there, OMC1 consists of dense fragments of gas and dust, in
part situated within the HII region, which itself constitutes the
visible Orion Nebula. The interest in OMC1 and the surrounding zone
stems from widespread ongoing star-formation in this region (e.g. Luhman
et al. 2000), exemplified through the presence of many protostars,
outflows and larger scale flows, for example the explosive
Becklin-Neugebauer-IRc2 complex (BN-IRc2) in OMC1 (see O'Dell 2001;
Allen\&Burton 1993; Doi et al. 2002; O'Dell\&Doi 2003). 

\begin{figure*} 
\centering
\includegraphics[width=16cm]{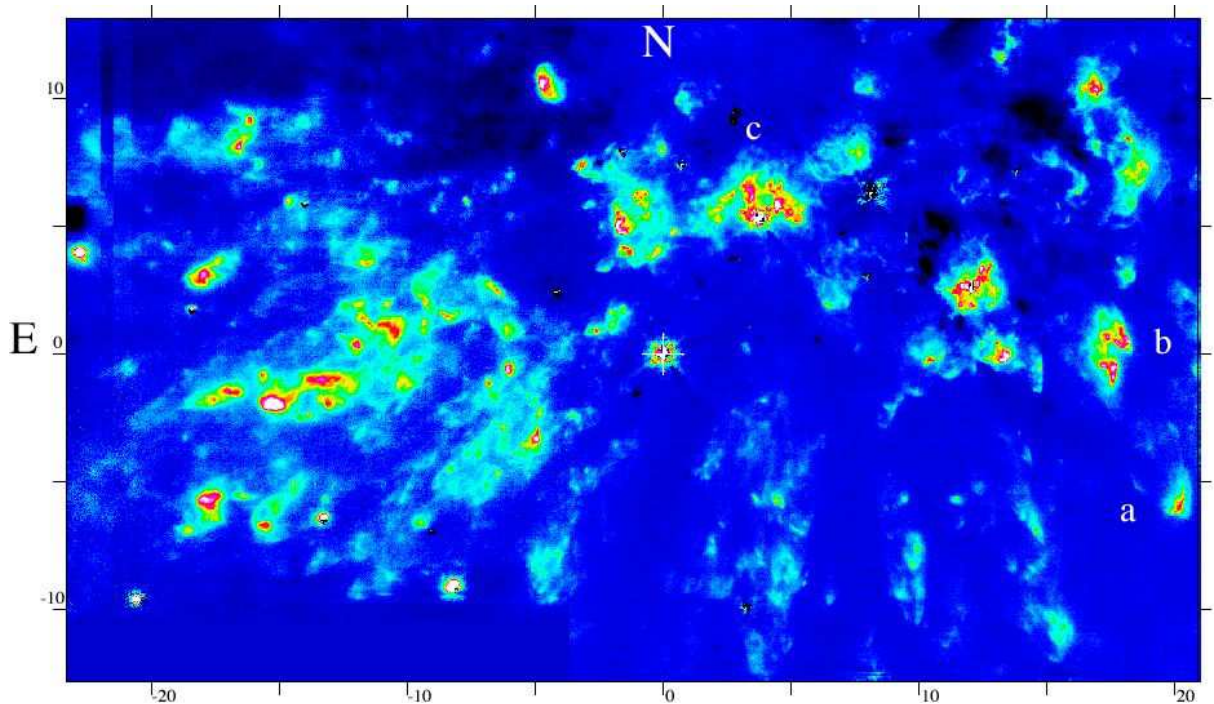}
\caption{H$_2$ emission in the 1-0 S(1) line at 2.121$\mu$m, with continuum 
emission subtracted. Image size: 44.5"x26". 
The  cross indicates the position of the star TCC016 
(05$^{\mathrm{h}}$35$^{\mathrm{m}}$14\fs91,
  -05\degr22'39\farcs31 (J2000)) and a,b,c identify objects shown in more detail 
in subsequents. The SE and ESE frames, referred to in the text, are 
respectively the regions to the west and to the east of TCC0016.}
\label{s1total}
\end{figure*}

In the present work we use the VLT-UT4 with NACO, an association of
the new adaptive optics system NAOS (Nasmyth Adaptive Optics System)
and  the CONICA (COud\'e Near IR Camera) infrared array camera (Lenzen
et al. 1998; Rousset et al. 2000; Brandner et al. 2002). This yields, in
part of the full image, a spatial resolution two to three times higher
than that achieved previously with the HST (Stolovy et al. 1998; Schultz
et al. 1998) or with 4m-class telescopes using adaptive optics e.g. the
Canada-France-Hawaii Telescope (V2001; G 2003; K2003).

Our present data allow us to address the issue of the scale sizes of
clumps of excited H$_2$ in OMC1 with significantly greater precision and
over a region within OMC1 roughly an order of magnitude larger in the
plane of the sky than in earlier work (V2001). In this connection, V2001
showed that H$_2$ emitting regions did not exhibit fractal structure but
showed a preferred scale of between 3 and 4 10$^{- 3}$ pc. Here we
corroborate this finding and also identify a second preferred scale size
of about half this value, in a neighbouring region of OMC1, to the west
of that studied in V2001.

\section {Observations and data reduction}

Observations of OMC1 were performed with the VLT UT4 during guaranteed
NACO time on the 21$^{st}$ of November 2002, for about 4.5 hours in
total. The seeing was variable during this night ranging from 0.65" to
1.13". Two regions have been observed located south-east (SE frame) and
east-south-east (ESE frame) from BN and IRc2, with the star TCC016 (Fig.
1) common to both frames. The IR wavefront sensor (WFS) was used for the
SE frame using BN (m$_K$$\sim$8) as the adaptive optics (AO) reference
star. The average seeing was $\sim$0.85" and the resulting FWHM of the  PSF measured
on several stars in the SE field was $\sim$0.08". The ESE frame was
recorded using the visible WFS locked on the star TCC016 (m$_V$$\sim$14)
which, combined with an average seeing of $\sim$1.03", led to a measured
FWHM of the PSF of $\sim$0.12". The Strehl ratio was clearly higher in the SE frame,
giving nearly full correction by the AO system to the diffraction limit
of $\sim$0.06". Using camera mode S27, the field of view was
27.6"$\times$27.6" and the pixel scale was 0.027", a size sufficient to
satisfy the Nyquist sampling criterion. 0.027" corresponds to 12 AU at
the distance of Orion.

Filters were used at 2.121$\pm$0.011$\mu$m (NB212), which contains the
H$_2$ S(1) v=1-0 line, IB224, at 2.24$\pm$0.03$\mu$m and IB227 at
2.27$\pm$0.03$\mu$m. IB224 includes the H$_2$ 1-0 S(0) line at
2.223$\mu$m and the H$_2$ 2-1 S(1) line at 2.247$\mu$m, as well as
[FeII] lines at 2.218 and 2.242$\mu$m. IB227 includes the H$_2$ 2-1 S(1)
line and the [FeII] line at 2.242$\mu$m. The  AutoJitter mode
was used, that is, at each exposure, the telescope  moves
according to a random pattern in a 6"$\times$6"
box. Cross-correlation was used to recenter the images at $\approx$0.15 pixel.
 For the ESE frame, an empty
background sky has been recorded for the purpose of sky subtraction. For
the other less crowded SE frame, the background sky has been created by
performing a median filtering of the set of 32 randomly jittered
individual frames. The total exposure time on object for each filter was
800 seconds for the SE frame and 1500 seconds for the ESE frame, with
exposures of 10 seconds for each data acquisition.

To obtain an image in the H$_2$ v=1-0 S(1) line, the IB244 continuum 
image was subtracted from the NB212 image after division by a factor of ~2.5 as 
derived from measuring fluxes of a dozen of stars  in both filters.The resulting image is
shown in Fig.\ref{s1total}. The contribution of v=1-0 S(0) and v=2-1
S(1) in IB224, known to be small from spectroscopy, was ignored in image
processing.
By subtraction of the image in the IB227 filter from that in the IB224,
using a conversion factor obtained as before from stellar fluxes, 
it was possible to obtain an image in H$_2$ S(0) v=1-0,
less the weaker emission in v=2-1 S(1). The image in v=1-0 S(0) shows
essentially the same structure as that in Fig.\ref{s1total}, but is 4 to
5 times weaker. An example of data in v=1-0 S(0) is shown in
Fig.\ref{closeupb}c.

\begin{figure} 
\centering
\includegraphics[width=8cm]{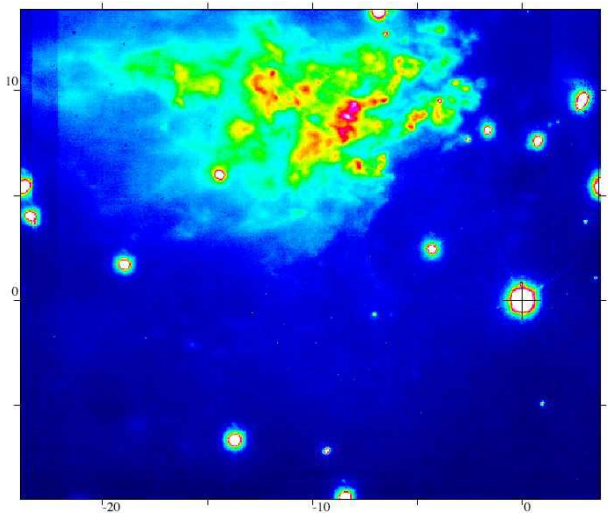}
\caption{Continuum emission at 2.27$\pm$0.03$\mu$m in the ESE field. Image size: 
27.6"x23.4". The position of TCC0016 is marked with a cross.}
\label{continuum}
\end{figure}

\begin{figure} 
\centering
\includegraphics[width=8cm]{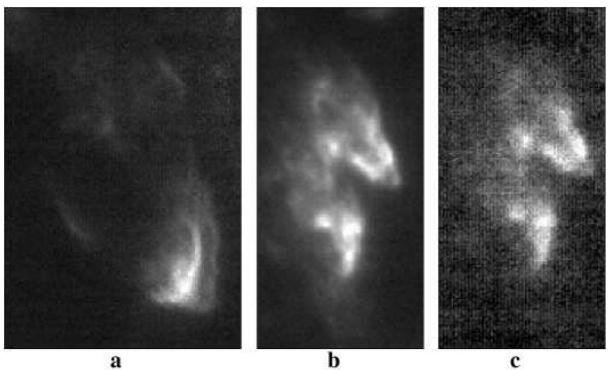}
\caption{a). A detailed view (3.5"x5.0") of object a (see Fig.\ref{s1total}) in 
v=1-0 S(1) H$_2$ emission at 2.121$\mu$m. b). of object b (2.5"x5.0") in the 
same line.  
c). Detail (2.5"x5.0") of object b in the 1-0 S(0) H$_2$ emission line at 
2.22$\mu$m.}
\label{closeupb}
\end{figure}

Strong continuum emission is seen for example in the northern part of
the ESE region and is shown in Fig.\ref{continuum}, which displays
emission in the IB227 filter. Data in Fig.\ref{continuum} also include
weak emission from v=2-1 S(1) and [Fe II] (Schultz et al.1999).
Continuum emission is also seen at this position in the data of Schultz
et al. 1999.

\begin{figure}
\centering
\includegraphics[width=8cm]{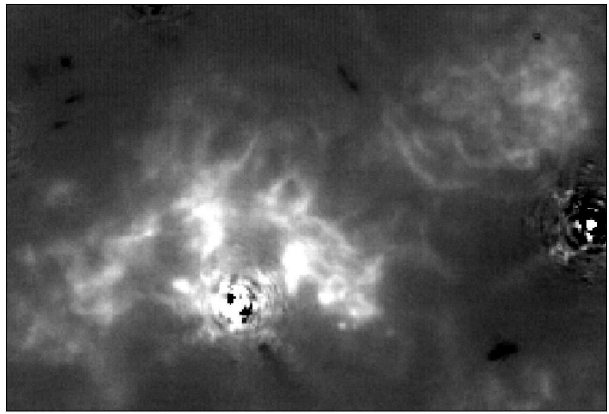}
\caption{A detailed view of object c (see Fig.\ref{s1total}) in
v=1-0 S(1) H$_2$ emission at 2.121$\mu$m. Image size: 7.5"x5.0".}
\label{closeupc}
\end{figure}

\section{Discussion of Results}

\subsection{Origin of the NIR emission}

H$_2$ emission in OMC1 arises from heating in J-type and C-type
(magnetic) shocks (e.g. Draine et al. 1983, Pineau des For\^ets et
al.1988, Smith\&Brand 1990, Kaufman\&Neufeld 1996a,b, Timmerman
1998, Wilgenbus et al. 2000, V2001, Le Bourlot et al. 2002, K2003, G2003) 
and from photon excitation in
photodissociation regions (PDRs) e.g. St\"orzer\&Hollenbach 1999,
Sternberg\&Dalgarno 1989, Black\&vanDishoeck 1987, Black\&Dalgarno 1976.

Shocks in the region observed arise from one or more large-scale
outflows, thought to originate from the BN-IRc2 region (O'Dell\&Doi
2003, Doi et al. 2002, O'Dell 2001 and references therein). Shocks in
the zone observed may also arise from local outflows associated with
protostars buried within emitting clumps of gas (G2003).

PDRs are generated through the action of the Trapezium stars, to the
south of the present region, of which the dominant contributor is
$\theta$$^1$Ori~C. This star generates far-UV radiation fields of
$>$10$^5$ times the standard interstellar field over much of OMC1, and
is in addition the chief contributor to the HII region in which dense
clumps in OMC1 may be bathed. The BN object is a young and massive
B-star (Gezari et al. 1998) and may also provide a source for PDR
excitation.

The relative importance of PDR and shock excitation in the strongest
emitting regions has been discussed in detail in K2003.
The conclusion is that in the subset of clumps studied in that work,
which are those to the east of TCC016 in the ESE image, C-type shocks
are the major contributor to H$_2$ excitation. However, towards the
fringes of bright clumps, J-type shock and PDR excitation by
$\theta$$^1$Ori~C take the place of C-type shock excitation, with shock
and PDR excitation contributing roughly equally. In all cases of very
bright emission, both in the cores of clumps and at their fringes,
the densities are high, exceeding several times 10$^7$
to $>$10$^8$cm$^{-3}$. 
Excited gas, formed in numerous, individually unresolved shocks,
rapidly accumulates into cold compressed zones and excited H$_2$ is the
very rapid progenitor of cold dense gas. Thus the structure of excited
H$_2$, discussed in section 3.2 below, provides a measure of the
structure of cold H$_2$ clumps, where the latter is the gas from which
stars will ultimately form.

\begin{figure}
\centering
\includegraphics[width=8cm]{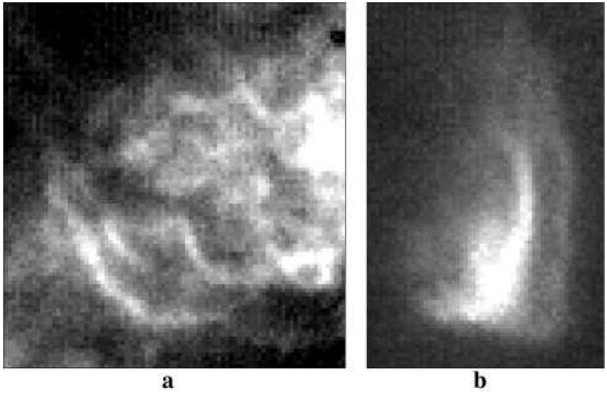}
\caption{a). Detail (1.56"x2.40") of the NW part of Fig.\ref{closeupc} and b). 
of Fig.\ref{closeupb}a (2.24"x2.40") showing narrow filamentary structure.}
\label{filament}
\end{figure}

In other regions, which are presumably less dense, the correspondingly
larger shock structure appears to be clearly resolved in H$_2$ emission.
This is illustrated by data in Fig.~\ref{closeupb}, \ref{closeupc} and
\ref{filament}. Numerous objects in these figures resemble bow shocks.
The general appearance of this figure suggests that the region may be
permeated by supersonic turbulence giving rise to shocks in all
directions. The emission may also in part be due to complex density
structure with PDR excitation. At all events, our NACO data allow us to
specify the width of the filamentary regions containing excited gas. For
example in Fig.\ref{filament}, showing the western part of object c
(Fig.\ref{closeupc}) and of object a (Fig.\ref{closeupb}) in greater
detail, widths of some filaments appear marginally resolved.
If these scales of $\approx$40--50 AU 
are shock widths, these data provide important constraints
on shock models (Le Bourlot et al. 2002, Wilgenbus et al. 2000), in
which parameters of density, magnetic field and shock speed determine
the shock width. If the features are PDRs, the widths of structures are
again valuable parameters, providing a good indicator of the gas density
(K2003, Lemaire et al. 1996, Sternberg\&Dalgarno 1989).
An additional constraint to shock or PDR models is provided by data for
the v=1-0 S(0) line, see Fig.\ref{closeupb}c. In future work, we will
pursue modeling of these regions, including constraints arising from
absolute brightness in the H$_2$ emission lines.

The origin of the diffuse continuum emission (plus [FeII] and some weak H$_2$
emission - see above), for example in Fig.\ref{continuum}, is most
likely reflected light from dust. In this connection, there is little
shock excitation where strong continuum emission is observed, as a
comparison of Fig.\ref{s1total} and \ref{continuum} reveals. Thus little
mechanical energy is being injected into this zone. Emission could arise
from far-UV photon heating of very small dust particles, as for example
in NGC7023 (Lemaire et al. 1996), with $\theta$$^1$Ori~C and BN as
sources in the present case. There is however strong circular
polarization in the K-band continuum in the zone shown in
Fig.\ref{continuum} (Chrysostomou et al. 2000). Those authors ascribed
the origin of circular polarization to scattering from oblate grains
oriented in magnetic fields. There is in fact striking spatial
correlation, lying within the observational resolution of the circular
polarization data, between the regions of strong circular polarization
reported in Chrysostomou et al. 2000 and the continuum emission seen in
Fig.\ref{continuum} and at other positions throughout the SE and ESE
images. If we include the linear polarization data of Geng 1993, then
effectively in all regions in which continuum emission is observed,
either circular or linear polarized K-band radiation may be found. This
lends strong support to the hypothesis that the observed diffuse emission in the
continuum is due to scattering from dust. According to Chrysostomou et
al. 2000, the source of illumination is not the Trapezium stars, but is
intrinsic to OMC1 through some unidentified source.

Weak [FeII] emission is present for example in the region shown in
Fig.\ref{continuum} (Schultz et al.1999) and most likely arises from
photoionization and excitation by $\theta$$^1$Ori~C in the HII plasma as
suggested in Schultz et al. 1999. In this model, [FeII] emission is
projected over the K-band continuum emission discussed above.

\subsection{Analysis of small scale structures}

V2001 gave the first clear indication that the size distribution of hot
H$_2$ clumps in Orion was not fractal, using data for a small region of
OMC1, 12.8"$\times$12.8", in the southern part of the ESE region of the
present work. In order to analyze the global nature of small scale
structure, V2001 used area-perimeter, Fourier, brightness histogram
(Blitz\&Williams 1997) and direct measurement analyses. All techniques
revealed that structure observed in v=1-0 S(1) H$_2$ emission was not
fractal and all yielded the same preferred scale size. Here we choose
the brightness histogram method for its simplicity and ease of
interpretation.

In brief, this method involves counting in the original image  the
number of pixels N(B$_n$), binned into some small normalized brightness
range, as a function of the pixel normalized brightness, B$_n$. B$_n$ is
defined as the value of brightness in any pixel divided by the value in
the brightest pixel in the image. The resolution of the image is then
degraded by boxcar averaging over some chosen number of pixels on a square and
a new set of N(B$_n$) and B$_n$ calculated. This process is repeated,
successively degrading the resolution of the image. Plots of N(B$_n$)/N$_{Tot}$
vs. B$_n$, where N$_{Tot}$ is the total number of pixels,
yield a set of curves which remain unchanged for an image with
no preferred scale, that is, a fractal image, as the resolution of the
image is degraded. By contrast, for an image with a preferred scale,
such plots should change in form as the resolution is degraded to the
point at which any preferred scale has been washed out. Beyond this
critical smoothing, the form of such plots should remain constant,
mimicking a fractal. The amplitude of the critical smoothing is a direct
measure of the preferred range of scale in the image.
In all cases stars were masked out in the images before the brightness
histogram analysis was performed.

\begin{figure} 
\centering
\includegraphics[width=8.3cm]{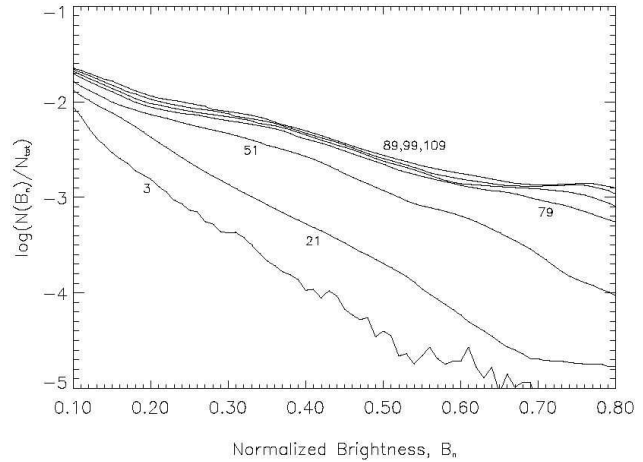}
\caption{Brightness histogram analysis of the H$_2$ v=1-0 S(1) emission at 
2.121$\mu$m in the ESE region. Figures besides the curves shown represent the 
number of pixels on the square over which the data have been spatially 
averaged.}
\label{blitz}
\end{figure}

A brightness histogram analysis for the ESE image in H$_2$ is shown in
Fig.\ref{blitz}. We estimate that the preferred scale for regions of
excited H$_2$ is encountered when averaging is performed over
90$\pm$5$\times$90$\pm$5 pixels, that is, at a scale of 2.4$\pm$0.14"
(1075 AU). An exactly similar analysis on the SE region, that is, to the
west of TCC016 in Fig.\ref{s1total}, yields a preferred scale of
1.2$\pm$0.14" (540 AU). If we choose a region which samples
approximately equally both the ESE and SE regions and perform a
brightness histogram analysis, we encounter two scales present together,
of 2.4" and 1.2". We conclude therefore that the ESE region contains
clumps of material of about twice the scale of the material in the SE
region, with a well-defined demarcation between them. The same analysis
has been applied to the image in the continuum, shown in
Fig.\ref{continuum}. This yields a scale of 2.15$\pm$0.28". This
indicates that dust, and by implication, cold unexcited gas in this
region is clumped on the same scale as in the shocked southern region in
ESE. The scale size, if one is present, of continuum emission in the SE region
could not be determined using the technique described here, but is larger
than 2.7 arcseconds. Data covering a more extensive field in OMC1 may
yield information in the future.

The scale-size identified in V2001, for a subset of the ESE region, was
2.0$\pm$0.1", in substantial agreement with the present work. As
described in detail in V2001, the brightness of emission of H$_2$ may be
used in conjunction with C-type magnetic shock models to show that in
bright emissions regions the dense gas, swept up by the shock, will
achieve densities of several times 10$^7$ cm$^{-3}$, as the post-shock
gas cools to 10K. Related data for the v=2-1 S(1) emission line,
discussed in K2003, show that densities may be locally
even higher, exceeding 10$^8$ cm$^{-3}$ in some regions. Our present
estimates of clump size add weight to the conclusion in V2001, based
upon Jeans length considerations, that sufficient gas has accumulated
through shock induced compression to form gravitationally unstable
clumps which are potential sites of star formation, in the ESE region.
This shock-induced mechanism would not appear to be presently effective
in hastening star-formation in the SE region, since we find that the
clump masses are typically an order of magnitude smaller than in ESE and
comprise no more than 0.01 to 0.02 M$_{\odot}$ if we assume densities
of $< 10^{8} \mbox{cm}^{-3}$ given the observed size of 1.2".
 In connection with the
interpretation of our data, results in G2003 appear to
show that protostars may already have formed in certain of the zones
very bright in H$_2$ emission. In some regions we may therefore be
observing the aftermath of very recent star formation rather than the
period just prior to gravitational collapse. 

\begin{acknowledgements}
DF and MG  acknowledge the support of the Aarhus Centre for
Atomic Physics, funded by the Danish Basic Research Foundation.
JLL and DR  acknowledge support of the french PCMI
program, funded by CNRS. 
\end{acknowledgements}


\begin{thebibliography} {}
\footnotesize
\bibitem[1993]{Allen} Allen,D.A.\&Burton,M.G. 1993, Nature 363, 54

\bibitem[2000]{Bally} Bally,J., O'Dell,C.R., McCaughrean,M.J. 2000, AJ, 119, 2919

\bibitem[1976]{Black} Black,J.H., Dalgarno,A., 1976, ApJ 203, 132

\bibitem [1987]{Black87} Black,J.H., van Dishoeck,E.F., 1987, ApJ 322, 412

\bibitem [1997]{Blitz} Blitz, L., and Williams, J.P., 1997, ApJ 488, L145 

\bibitem [1995]{Brand} Brand,P.W.J.L. 1995, ApJSS 233, 27

\bibitem [2002]{Brandner} Brandner, W., Rousset, G., Lenzen, R., et al. 2002, The ESO 
Messenger, 107, 1

\bibitem [1998]{Chen} Chen,H., Bally,J., O'Dell,C.R. et al. 1998, ApJ 492, L173
 
\bibitem [1997]{Chrysostomou} Chrysostomou ,A., Burton,M.G., Axon,D.J., Brand,P.W.J.L., Hough,J.H., Bland-Hawthorn,J., Geballe,T.R. 1997, MNRAS 289, 605 

\bibitem [2000]{Chrysostomou00} Chrysostomou,A., Glendhill,T.M., M\'enard,F., Hough,J.H., Tamura,M., Bailey, J. 2000, MNRAS 312, 103

\bibitem [1983]{Draine} Draine,B.T., Roberge,W.G., Dalgarno,A. 1983, ApJ 264, 485

\bibitem [2001]{Ferland} Ferland,G.J., 2001, PASP, 113, 41

\bibitem [2000]{Geng} Geng F., PhD Thesis, University of Tokyo. Data are shown in 
Chrysostomou et al. 2000.

\bibitem [1998]{Gezari} Gezari,D.Y., Backman,D.E.,Werner,M.W. 1998, ApJ, 509, 283

\bibitem [2003]{Gustafsson} Gustafsson,M., Kristensen,L.E., Cl\'enet,Y., Field,D., Lemaire, J.L., Pineau des For\^ets,G., Rouan,D., Le Coarer,E. {\it submitted to A\&A} 2003

\bibitem [1996]{Kaufman} Kaufman,M.J., Neufeld,D.A., 1996, ApJ, 456, 250  and 611

\bibitem [2003]{2003} Kristensen, L., Gustafsson, M., Field, D., Callejo, G., Lemaire,J.L., Vannier,L., Pineau des For\^ets,G., {\it submitted to A\&A} 2003


\bibitem [2002]{Le Bourlot} Le Bourlot,J., Pineau des For\^ets,G., Flower,D.R. Cabrit,S., 2002, MNRAS, 332, 985

\bibitem [2000]{Lee} Lee,J.-K.\&Burton,M.G. 2000, MNRAS 315,11

\bibitem [1996]{Lemaire} Lemaire, J.L., Field, D., Gerin, M., Leach, S., Pineau des 
For\^ets,G., Rostas, F., Rouan, D., 1996, A\&A 308 895

\bibitem [1998]{Lenzen} Lenzen, R., Hofmann, R., Bizenberger, P., Tusche, A., et al. 1998, \procspie, 3354, 606

\bibitem [1997]{McCaughrean} McCaughrean,M.J., MacLow,M.M. 1997, AJ 113, 391

\bibitem [2001]{O'Dell} O'Dell,C.R. 2001, Ann.Rev.Astron.Astrophys., 39, 99

\bibitem [2003]{O'Dell03} O'Dell,C.R.\&Doi,T. 2003, AJ 125, 277

\bibitem [1988]{Pineau } Pineau des For\^ets,G., Flower,D.R.,Dalgarno,A. 1988 MNRAS, 235, 621

\bibitem [2000]{Rousset} Rousset, G., Lacombe, F., Puget, P., et al. 2000, \procspie, 4007, 72 

\bibitem [1999]{Salas} Salas,L., Rosado,M., Cruz-Gonzales,I., Gutierrez,L., Valdez,J. et al. 1999 ApJ, 511, 822

\bibitem [1997]{Schild} Schild,H.,Miller,S.\&Tennyson,J. 1997, A\&A 318, 608   

\bibitem [1999]{Schultz} Schultz,A.S.B., Colgan,S.W.J., Erickson,E.F., Kaufman,M.J., 
Hollenbach,D.J., O'Dell,C.R., Young,E.T., Chen,H. 1999 ApJ 511 282

\bibitem [1990]{Smith} Smith,M.D., Brand,P.W.J.L. 1990, MNRAS, 242, 495

\bibitem [1989]{Sternberg} Sternberg,A., Dalgarno,A. 1989, ApJ 338,197


\bibitem [1998]{Stolovy} Stolovy,S.R., Burton,M.G., Erikson, E., Kaufman, M.J. et al. 1998, ApJ 492, L151

\bibitem [1999]{Storzer} St\"orzer,H., Hollenbach,D.J. 1999, ApJ, 515, 669

\bibitem [1999]{Tedds} Tedds,J.A., Brand,P.W.J.L., Burton,M.G. 1999, MNRAS 307, 337


\bibitem [1998]{Timmerman} Timmerman,R. 1998, ApJ, 498, 246

\bibitem [2001]{Vannier}  Vannier, L., Lemaire, J.L., Field, D., Pineau des For\^ets, G.,
Pijpers, F.P. \& Rouan, D. 2001, A\&A 366, 651

\bibitem [2000]{Wilgenbus} Wilgenbus,D., Cabrit,S., Pineau des For\^ets,G., Flower,D.R. 2000, 
A\&A, 356, 1010


\end{thebibliography}
\end{document}